\documentstyle[12pt]{article}
\newcommand{\be}{\begin{equation}}
\newcommand{\ee}{\end{equation}}
\newcommand{\ul}{\underline}
\begin{document}
\def\theequation{\arabic{section}.\arabic{equation}}
\begin{titlepage}
\title{The creation of multiple images \\
by a gravitational wave}
\author{Valerio Faraoni \\ \\ {\small \it Department of Physics and Astronomy,
University of Victoria, P.O. Box 3055}\\
{\small \it  Victoria, B.C. Canada V8W 3P6}}
\date{}     \maketitle      \thispagestyle{empty}
\vfill
\begin{abstract} 
We describe gravitational lensing by a gravitational wave, in the regime in
which multiple images of a light source are created. We adapt 
the vector formalism employed for ordinary gravitational 
lenses to the case of a non--stationary 
spacetime, and we derive an approximate condition for multiple imaging. 
It is shown that certain astrophysical sources of gravitational waves satisfy
this condition. 
\end{abstract}      \vspace*{1truecm}   \begin{center}
To appear in Proceedings of the Pacific Conference on Gravitation and
Cosmology, Seoul, Korea, 1--6 February 1996.
\end{center}    \end{titlepage} \clearpage   \setcounter{page}{1}

We study gravitational lensing with a gravitational wave acting as the lens. 
It is well--known that exact gravitational waves deflect light, and 
light propagation through linearized (cosmological) 
gravitational waves outside the laboratory
has also been studied \cite{gw}, expecially in 
conjunction with their frequency shift 
effect on the photons of the cosmic microwave background. It has also been
suggested that gravitational waves may create multiple images of distant light
sources \cite{BBW}. We concentrate on the latter aspect of lensing by
gravitational waves, for two reasons: {\em i)}~the possibility of multiple
images is associated to strong amplifications of the light source, which 
makes easier to detect gravitational waves; {\em ii)}~a previous analysis 
\cite{BBVarenna} concluded that the amplification 
is negligible. This conclusion was 
based on the
Raychaudhuri's equation, and it is not valid when multiple images are involved.

\section{The vector formalism for lensing gravitational waves}

We consider a gravitational wave localized in a region of space between
a light source and an observer. The spacetime metric is
$ g_{\mu \nu }=\eta_{\mu \nu}+h_{\mu \nu} $, where $ \eta_{\mu \nu} $
is the Minkowski metric and $ |h_{\mu \nu}| \ll 1 $.
Let us consider a light ray whose unperturbed path is parallel to the $ z
$-axis, with 4-momentum
$ p^{\mu}=p^{\mu}_{(0)}+\delta p^{\mu }=(1+\delta p^{0},\delta
p^{1},\delta p^{2},1+\delta p^{3}) $,
where $\delta p^{\mu } $ are small deflections. We work in the geometric
optics approximation, which holds if $ \lambda_{g.w.} >>\lambda_{e.m.} $
and $ \lambda_{e.m.} >{\lambda_{g.w.}}^2 /D_{L} $ (where $ \lambda_{g.w.} $
and $ \lambda_{e.m.} $ are the wavelengths of
the gravitational wave and of the photon, respectively, and $ D_{L} $ is
the observer-lens distance).  
We set the geometry as customary in gravitational lens
theory, using lens and source planes with coordinates $ (x,y) $ and
$ (s_{x},s_{y}) $ respectively. The action of the
lens is described by the plane-to-plane mapping $ x^{A} \longmapsto
s^{A} $ ($ A=1,2 $), with $ \ul{s} $ given by the ``lens
equation''\setcounter{equation}{0}
\be    \label{lensequation}
s^{A}=x^{A}+\frac{D_{L}D_{LS}}{D_{S}}\,\, \delta p^{A}( \ul{x} ) \ee
($ D_{LS} $ and $ D_{S} $ are the lens-source and the observer-source
distances respectively). 

The map described by Eq.~(\ref{lensequation}) has the Jacobian matrix
\begin{equation}
J\left( \begin{array}{c}
\underline{s} \\ \underline{x}
\end{array}   \right) =\left( \frac{\partial s^{A}}{\partial x^{B}} \right)
=\left( \begin{array}{cc}
1+D \,\partial_{x}( \delta p^{x})  & D\, \partial_{y}( \delta p^{x}) \\
D\,\partial_{x}( \delta p^{y})   & 1+D\,\partial_{y} ( \delta p^{y})
\end{array}  \right) \; ,
\end{equation}
where $ D\equiv D_{L}D_{LS}/D_{S} $. The inverse matrix $ A=J^{-1} $
represents the {\em amplification tensor}, while its determinant $ {\cal
A}=\mbox{Det}(J)^{-1} $ is the (scalar) {\em amplification}. Since the 
surface
brightness is conserved during lensing whenever geometric optics holds
and
\begin{equation}
 {\cal A}=\frac{area\,\,\,\,of
\,\,\,\,an\,\,\,\,infinitesimal\,\,\,\,region\,\,\,\,in\,\,\,\,the\,\,\,\,
lens\,\,\,\,plane}{area\,\,\,\,of\,\,\,\,the\,\,\,\,corresponding\,\,\,\,
region\,\,\,\,in\,\,\,\,the\,\,\,\,source\,\,\,\,plane} \:\:,
\end{equation}
$\cal A $ has also the meaning of the ratio of light intensities with and
without the lens. A small circular source will be imaged into
a small ellipse whose eccentricity $ e $ is given by the ratio of the
eigenvalues $ e_{\pm} $ of $ A $:
\begin{equation}
 (1-e^{2})^{1/2}=\left| \frac{e_{+}}{e_{-}}\right| \;.
\end{equation}
The vanishing of the Jacobian $ \mbox{Det}(J) $ indicates the failure of
invertibility of the map~(\ref{lensequation}). Therefore, {\em the 
occurrence of multiple images is
signalled by the vanishing of} Det$( J)$, and this is the condition that we
study in the following. 

The Jacobian determinant Det($J$) can be written as follows:
\begin{eqnarray}
& & \mbox{Det}(J)=1+\sqrt{f( \alpha )}\,D_{S}\,
\frac{\partial ( \delta p^{A})}{\partial x^{A}}+
f( \alpha )D_{S}^{2} \left[
\partial_{x}( \delta p^{x}) \cdot \partial_{y}( \delta p^{y})-
\partial_{y}( \delta p^{x}) \cdot \partial_{x}( \delta p^{y}) \right]
\equiv  \nonumber \\
& & \equiv 1+J_{1}+J_{2}     \;,
\label{determinant}
\end{eqnarray}
where $ \alpha \equiv D_{LS}/D_{S}  $ and
$ f( \alpha )=\alpha^{2}(1-\alpha )^{2} $ is symmetric
about $ \alpha =1/2 $, where it assumes its maximum value $ 1/16 $. The
deflection $ \delta p^\mu $, computed from the equation of null
geodesics, is
\be \label{null}
\delta p^{A}=\frac{1}{2} \,\int_{Source}^{Observer} dz\,\,\left(
{h_{00}}+2{h_{03}}+{h_{33}} \right)^{,A} +O(h^2) \; . \ee
Moreover, $ \partial_A ( \delta p^A)=0$ to first
order and $ J_1 \sim h^2 D/P \ll J_2 \sim  h^2 (D/P)^2$ for large values 
of $ D/P $. Thus, in order to have multiple
imaging, it must be $ J_2 <0 $ and $
f( \alpha )\left[ D_{S}\, \partial_{A}( \delta p^{B}) \right] ^{2}\approx 1
$. For ordinary gravitational lenses, the probability of lensing of a distant
source is maximum when the lens is halfway between the source and
the observer, hence we take $ f( \alpha ) $ in the range
$ \frac{1}{100} $ -- $\frac{1}{16} $, near its maximum. Then, in order to
have multiple imaging, it must be 
\be  \label{roughcondition}
\frac{h}{P}\geq {\cal S}_{c} \; ,  
\ee
where $ P $ is the period of the gravitational wave,
${\cal S}_{c}\equiv (4-10) \, c/D $. The approximate condition for multiple
imaging~(\ref{roughcondition})
involves the ``strength'' $ h $ and the ``size'' $ P $ of gravitational waves,
and the geometry of the problem (through $ D $). (\ref{roughcondition})
is somewhat analogous to the well-known condition for multiple imaging by
ordinary gravitational lenses, $ \Sigma \geq
\Sigma_{c}\equiv \frac{c^{2}}{4\pi G}\cdot \frac{D_S}{D_L D_{LS}} $ .

\section{Lensing by gravitational waves and the Fermat principle}
 
The vector formalism for ordinary
gravitational lenses requires the stationarity of the lens
potential; clearly this hypothesis is not satisfied in our case, and a new
approach is needed. We substitute the true photon path in
the 3-dimensional space with a zig--zag path composed of two straight lines
from the source to the lens, and from the lens to the observer.
By applying a new version of the Fermat principle valid for
non(conformally)--stationary spacetimes \cite{Perlick}, it was shown in 
Ref.~\cite{meApJ} that the zig--zag 
paths are extrema of the travel time functional\setcounter{equation}{0}
\be \tilde{t}=\mbox{constant} \, + \frac{1}{2c}\cdot \left[ \frac{D_S}{D_L
D_{LS}}\, ( {\underline{x}}-{\underline{s}})^{2}+
\int_{S}^{O} dz \, (h_{00}+2h_{03}+h_{33} )  \right] \;.  \ee
The lens equation~(\ref{lensequation}) and the deflection (\ref{null}) 
are obtained by requiring stationarity of this functional: 
$ \underline{\nabla}_{\underline{x}}\tilde{t}=0$. 
Therefore, the conclusions of the previous section are justified in the
context of a rigorous formalism.

\section{Comparison between a gravitational wave and an ordinary gravitational
lens}

We compare the action of a gravitational wave with that
of an ordinary gravitational lens. The latter is a mass distribution described
by a Newtonian potential $ \Phi $ (satisfying the Poisson equation $
\nabla^{2} \Phi =4\pi \rho $, where $ \rho $ is the lens mass density). 
The plane--to--plane map describing the lens action is given by the lens
equation and the Jacobian matrix can be written as\setcounter{equation}{0}
\begin{equation}
J=\left( \begin{array}{cc}
1-\chi -\Lambda    &      -\mu \\
-\mu               &      1-\chi +\Lambda
\end{array}   \right) \;,
\end{equation}
with
\begin{eqnarray}
\chi & \equiv & \frac{\Sigma}{\Sigma_{c}} \;, \\
\Lambda & \equiv & \frac{D}{c^{2}} \int_{-\infty}^{+\infty} dl\, \left(
\frac{\partial^{2} \Phi }{\partial x^{2}}-
\frac{\partial^{2} \Phi }{\partial y^{2}}   \right)  \;,  \\
\mu & \equiv & \frac{D}{c^{2}} \int_{-\infty }^{+\infty} dl \,\,
\frac{\partial^{2}\Phi }{\partial x\partial y} \; ,
\end{eqnarray}
where $ D $ and $\Sigma_c$ have been defined in the previous sections and 
\begin{equation}
\Sigma \equiv \int_{-\infty}^{+\infty}dl\, \rho  \; .
\end{equation}
The Jacobian determinant is given by
\begin{equation}
\mbox{Det}(J)=(1-\chi )^{2}-( \Lambda^{2}+\mu^{2})  \;.
\end{equation}
The {\em convergence} $ \chi $ describes the action of matter, while $
\Lambda $ and $ \mu $ describe the action of shear.
For a lensing gravitational wave one obtains 
\begin{equation}
J_{gw}=\left(   \begin{array}{cc}
1-\Lambda_{1}  &    -\mu_{1}   \\
-\mu_{2}       &    1-\Lambda_{2}
\end{array}   \right) \;.
\end{equation}
where, to first order, one has
\begin{eqnarray}  \label{landa1}
\Lambda_{1} & \equiv & -\frac{D}{2} \int_{S}^{O} d\lambda \,\,
{{h_{\lambda \lambda }}^{,x}}_{,x} \;, \\
\label{landa2}
\Lambda_{2} & \equiv & -\frac{D}{2}\int_{S}^{O} d\lambda \,\,{{h_{\lambda
\lambda}}^{,y}}_{,y} \;,  \\
\label{mu1}
\mu_{1} & \equiv & -\frac{D}{2} \int_{S}^{O} d\lambda \,\,
{{h_{\lambda \lambda }}^{,x}}_{,y} \;,  \\
\label{mu2}
\mu_{2} & \equiv & -\frac{D}{2}\int_{S}^{O} d\lambda \,\,
{{h_{\lambda \lambda}}^{,y}}_{,x} \;,
\end{eqnarray}
where the integrals are computed along the photon's path from the source
to the observer. To first order, 
$ \Lambda_{2}=-\Lambda_{1}\equiv -\Lambda_{gw}
$ and $ \mu_{1}=\mu_{2}\equiv \mu_{gw} $. To the lowest order, we have 
\begin{equation}
  J_{gw}=\left(   \begin{array}{cc}
1-\Lambda_{gw}          &       -\mu_{gw}       \\
-\mu_{gw}               &       1+\Lambda_{gw}   \end{array}   \right) \; .
\end{equation}
The convergence
term is absent and hence the lens action is due only to the shear. This
result was derived in Ref.~\cite{BBVarenna} using the Raychaudhuri equation 
and the optical scalars formalism, and is now recovered in the 
vector formalism.

The deflection angle does not depend on
the frequency of the light: gravitational waves are {\em
achromatic} lenses, like ordinary gravitational lenses. However, while the
latters do not shift the frequency of the photons propagating through them,
lensing gravitational waves do. 
In addition, gravitational waves do not rotate the polarization plane of 
the electromagnetic field, to first order \cite{meAA}. In this 
aspect, they behave like ordinary gravitational lenses. 

\section{Order of magnitude estimates}

In order to apply the previous theory, and for the multiple images to be
detectable, the following conditions must be satified:
\begin{enumerate}
\item the scale of separation $ \delta \approx h $ between different images
must not be smaller than $ 10^{-3} $ arcseconds, which gives
\setcounter{equation}{0}
\be   \label{1}
h\geq 5\cdot 10^{-9}  \ee
at the deflection place;
\item the impact parameter $ r $ must satisfy
\be        \label{2}
r > \lambda_{g.w.} \; ;                             \ee
\item the lens must not be exceptionally rare;
\item the period of the lensing
gravitational wave must not be too short (let us say $ P <
10^{8} $~s), in order to appreciate variability of the images.
\end{enumerate}
Multiple imaging by gravitational waves is possible, and it occurs when 
the waves satisfy
the approximate inequality~(\ref{roughcondition}). We examine the validity of
this condition for the most common astrophysical
sources of gravitational radiation.\\
\noindent {\em Stellar core collapse}: using the values of $ h $ and $ P $
predicted in studies of collapsing homogeneous ellipsoids, 
(\ref{1}) and~(\ref{2}) allow only a rather narrow range
for the impact parameter $ r $, for which $ Dh/P \sim 10 $ if
$ D \sim 6 \cdot 10^{15} $ cm. The late phase when the ellipsoid has settled
down as a rapidly rotating neutron star does not give appreciable
lensing. Asymmetry due to the core's bouncing gives $ Dh/P \sim 10
$ if $ D \sim 10^{12} $ cm, $ r \sim 3\cdot 10^{7} $ cm, or if $ D \sim
6 \cdot 10^{16} $ cm, $ r \sim 2 \cdot 10^{12} $ cm. Extrapolating the
results obtained in studies of the
perturbations of pressureless spherical collapse leading to
the formation of a black hole, one obtains $ Dh/P \sim 10 $ if
$ D \sim 3\cdot 10^{16} $ cm (with a rather narrow range
of values of $ r $).\\
\noindent {\em Final decay of a neutron star/neutron star binary}: 
rough estimates for
the final decay of a binary system composed of two
neutron stars give $ Dh/P \sim 10 $ if
$ D\sim 3\cdot 10^{16} $ cm, $ r \sim 5\cdot 10^{10} $ cm, or if
$ D\sim 6\cdot 10^{7} $ cm, $ r \sim 6\cdot 10^{7} $ cm.\\
\noindent {\em Black hole collisions}: if two black holes collide 
with enough angular momentum to go into
orbit before coalescing, one has $ Dh/P \sim 10 $ if
$ D \sim 3\cdot 10^{17} $ cm, $ r\sim 3\cdot 10^{10} $ cm, or if
$ D\sim 6\cdot 10^{19} $ cm, $ r\sim 10^{12} $ cm.\\
\noindent {\em The binary pulsar}: the binary pulsar PSR 1913+16 is believed
to radiate gravitational waves in a continuous way with amplitude given,
in order of magnitude, by $
h \sim \ddot{Q}/r \sim Ma^{2}\omega^{2}/r $,
where $ Q $ and $ M $ are the quadrupole moment and the mass of the pulsar,
and $ a $ is the semimajor axis of the binary
system. Conditions~(\ref{1}) and~(\ref{2}) are incompatible, hence 
multiple imaging is not possible in this case.\\
\noindent {\em The gravitational wave background}: one has
$ Dh/P \approx \sqrt{\Omega_{g.w.}} D/R $, where
$ R $ is the radius of the universe and $ \Omega_{g.w.} $ is the
density of gravitational waves (in units of the critical
density). Upper bounds on $ \Omega_{g.w.} $ give $ \Omega_{g.w.} <1 $ for
all frequencies, and $ \Omega_{g.w.}\ll 1 $ in many bands. Moreover, since 
$ D/R \ll 1 $, also $ Dh/P \ll 1 $, and 
multiple imaging by the gravitational wave background is, on average,
impossible. 

As a conclusion, the creation of multiple images by a gravitational wave 
is possible and it is expected to occur. There are high amplification
events associated to multiple images, in contrast with previous conclusions
based on an inadequate formalism. The details of this work, and the study 
of the 
probability of observing multiple images and strong amplifications, and the 
details of the phenomenon will be the subject of a future publication
\cite{submitted}.

\section*{Acknowledgments}

The author is indebted to Prof. B. Bertotti for suggesting the
possibility of lensing by gravitational waves and the 
approach used in this paper, and to Prof. G.F.R. Ellis for stimulating
discussions. 


\clearpage
\begin{thebibliography}{99}

\bibitem{gw} D.M. Zipoy 1966, {\em Phys. Rev.} {\bf 142}, 825; G. Dautcourt
1975, {\em Astron. Astrophys.} {\bf 38}, 344; R. Fakir 1993, {\em Astrophys.
J.} {\bf 418}, 202; 1994, {\em Astrophys. J.} {\bf 426}, 74; 1994, {\em 
Phys. Rev. D} {\bf 50}, 3795.

\bibitem{BBW} B. Bertotti, private communication; J.A. Wheeler 1961, in {\em 
Rendiconti della Societ\`a Italiana di Fisica, 11th Course of the Varenna 
Summer School} (Academic Press, New York); A. Labeyrie 1993, {\em 
Astron. Astrophys.} {\bf 268}, 823.

\bibitem{BBVarenna} B. Bertotti 1971, in {\em General Relativity and
Cosmology}, R.K. Sachs ed. (Academic Press, New York).

\bibitem{Perlick} V. Perlick 1990, {\em Class. Quant. Grav.} {\bf 7}, 1319.

\bibitem{meApJ} V. Faraoni 1992, {\em Astrophys. J.} {\bf 398}, 425.

\bibitem{meAA} V. Faraoni 1993, {\em Astron. Astrophys.} {\bf 272}, 385.

\bibitem{submitted} V. Faraoni 1996 (submitted for publication).
\end{thebibliography}
\end{document}